\documentclass[prl,aps,twocolumn,showpacs,superscriptaddress,floatfix]{revtex4-1}

\usepackage{graphicx}
\usepackage{amsmath}

\newcommand{\ket}[1]{|#1\rangle}
\newcommand{\bra}[1]{\langle #1|}
\newcommand{\overlap}[2]{\langle #1|#2\rangle}
\newcommand{\xlap}[2]{X^{#1}_{#2}} 
\newcommand{\expect}[1]{\overline{#1}}

\newcommand{\nsys}{N_s}

\newcommand{\dosbath}{\nu_b}

\newcommand{\derivtau}[1]{\frac{\partial #1}{\partial\tau}}

\newcommand{\green}[1]{\frac{c_{#1}}{E^2_{#1}}}
\newcommand{\dtwopi}[1]{\frac{d{#1}}{2\pi}}

\newcommand{\egysys}{\varepsilon}
\newcommand{\egybath}{\epsilon}
\newcommand{\binit}{b_0}
\newcommand{\sinit}{s_0}
\newcommand{\rdmdiag}{\rho_{ss}}
\newcommand{\Vcouple}{V}
\newcommand{\Vbrown}{V_{\rm Br}}
\newcommand{\momM}{{\cal M}}
\newcommand{\momN}{{\cal N}}

\begin{document}

\title{Dynamics of Thermalization and Decoherence of a Nanoscale System}

\author{S.~Genway}\affiliation{School of Physics and Astronomy, The
  University of Nottingham, Nottingham NG7 2RD, United Kingdom}

\author{A.~F.~Ho}\affiliation{Department of Physics, Royal Holloway
  University of London, Egham, Surrey TW20 0EX, United Kingdom}

\author{D.~K.~K.~Lee}\affiliation{Blackett Laboratory, Imperial
  College London, London SW7 2AZ, United Kingdom}



\begin{abstract} 
  We study the decoherence and thermalization dynamics of a nanoscale
  system coupled nonperturbatively to a fully quantum-mechanical
  bath. The system is prepared out of equilibrium in a pure state of
  the complete system. We propose a random matrix model and show
  analytically that there are two robust temporal regimes in the
  approach of the system to equilibrium --- an initial Gaussian decay
  followed by an exponential tail, consistent with numerical results
  on small interacting lattices [S.~Genway, A.~F.~Ho and D.~K.~K.~Lee,
  Phys.~Rev.~Lett.~{\bf 105} 260402
  (2010)]. Furthermore, the system decays towards a Gibbs ensemble in
  accordance with the eigenstate thermalization hypothesis.

\end{abstract}

\maketitle

The origin of thermodynamics from a fully quantum-mechanical
description has been the subject of much recent
research~\cite{Polkovnikov2011, Yukalov2011, Cazalilla2010}.
Emergence of thermal behavior from the unitary evolution of a
wavefunction on a generic closed system can be studied using concepts
such as the eigenstate thermalization hypothesis~\cite{Deutsch1991,
  *Srednicki1994, *Rigol2008} (ETH) and canonical
typicality~\cite{Popescu2006, Goldstein2006, *Goldstein2010}. Local or
few-body observables in a closed nonintegrable system are expected to
`thermalize' at long times~\cite{Genway2012} in the sense that they
converge to a thermal Gibbs distribution. This has been studied with
various approaches~\cite{Tasaki1998, Reimann2008, *Reimann2010,
  Gemmer2001,*Mahler2005, *Gemmer2006, Brandao2008, Rigol2012} and for
myriad systems~\cite{Lesanovsky2010, Yuan2009, *Jin2010, Fine2009,
  *Ji2011, *Kolley2012, Kollar2011, Pal2010, Calabrese2011,
  Steinigeweg2013, Genway2010}. Recent interest has turned to
understanding the \emph{dynamics} of the relaxation to the thermal
state~\cite{Ates2012, *Ji2013, Genway2010, Miranda2010, Bartsch2009,
  Niemeyer2013, Mossel2010, Essler2012}.  In the canonical model, one
considers a composite of system and bath~\cite{Goldstein2006} and asks
how the system relaxes and decoheres~\cite{Zurek2003, *Zurek2009,
  Cucchietti2003} to reach a thermal state at long times.

In previous work~\cite{Genway2010}, we found numerically that a random
matrix model provided a generic description of thermalization dynamics
for (nonrandom) nanoscale Hubbard clusters. In this Letter, we provide
an analytical framework for this random matrix model. We derive
[Eqs.~(\ref{eq:rdmdiag_t}-\ref{eq:Lambda})] the relaxation
dynamics of a generic quantum system over the whole temporal range
from short to long times. We also confirm that the model produces a
thermal state at long times in accordance with ETH. Random matrix
methods have been employed to study nanoscale systems coupled to
different
environments~\cite{Mello1988,*Pereyra1991,Esposito2003,LebowitzPastur2004,David2011}.
However, they do not capture the full range of temporal behavior: 
there is no general account of the Gaussian decay towards
thermalization that has been established~\cite{Yuan2009,Genway2010}
numerically as a generic feature for the relaxation of local
observables in interacting systems.

We focus on a nanoscale system ($S$) with a discrete
energy spectrum embedded in a nonintegrable interacting bath ($B$)
with a quasicontinuous spectrum so that the average bath level
spacing $\Delta_B$ is much smaller than the system level spacings.
We will examine how the small system thermalizes with the bath
via the unitary evolution of the quantum-coherent
\emph{composite} system using a banded coupling model. Previous
authors studied a banded
coupling~\cite{Pereyra1991,LebowitzPastur2004} but were unable
to access the regime where we see Gaussian decay (see below).

{\it The model.}--- Suppose the system has $\nsys$ eigenstates $\ket{s}$
with energies $\egysys_s$ and the bath has eigenstates $\ket{b}$ of
energies $\egybath_b$.  The Hamiltonian for the composite system is
given by $H=H_0+V$:
\begin{equation}\label{eq:definemodel}
H =\sum_{sb} E_{sb}\ket{sb}\bra{sb} 
+ \sum_{ss'bb'}\ket{sb}\bra{sb}\Vcouple\ket{s'b'}\bra{s'b'}
\end{equation}
where $\ket{sb}\equiv \ket{s}\otimes\ket{b}$ are product states with
energies $E_{sb} =\egysys_s + \egybath_b$ for the decoupled system and
bath, and $\Vcouple$ couples the system and the bath.  The coupled
system will have an average level spacing of $\Delta=\Delta_B/\nsys$.
Analogous to the classic random matrix theory of nuclear matter, we
model the interacting bath with an energy spectrum that obeys
Wigner-Dyson statistics. Note that the randomness does not arise from
quenched disorder. We assume that the bath states $\ket{b}$ are
random vectors with no special spatial structure, \emph{e.g.}~no
spatial localization. This should be valid for generic interacting
quantum systems at energies away from strongly correlated states near
the bath ground state. The matrix elements of the coupling $\Vcouple$
in a basis involving these bath states should therefore also be
random. We use a banded random matrix of bandwidth $W$ and strength
$c$.  More precisely, the matrix element $\bra{sb}\Vcouple\ket{s'b'}$
is nonzero only if $|E_{sb}-E_{s'b'}| < W$, and each nonzero element
is a Gaussian random variable with zero average and a mean-square
value $\expect{|\bra{sb}\Vcouple\ket{s'b'}|^2} = c\Delta$. As we see
below, this scaling with the level spacing $\Delta$ is consistent with
a \emph{local} coupling between system and bath.

We can motivate this banded coupling model in the context of ultracold
atoms in optical lattices. A small cluster of sites (system) is
initially isolated from the rest of the lattice (bath) by a high
tunneling barrier. The coupling is introduced by lowering this barrier
to allow particles to hop between the cluster and the lattice.  This
particle exchange only couples bath states with an energy difference
of the order of the single-particle bandwith.  This produces a dense
banded matrix with bandwidth $W$ (see Fig.~20 of
Ref.~\cite{Genway2012} on the Hubbard model description of this
setup). This is the motivation of our banded coupling $\Vcouple$.  Our
scaling of the coupling with the level spacing $\Delta$ is also
motivated by the \emph{local} quench in this lattice example:
$\textrm{Tr}\, \Vcouple^2 \propto d NJ_h^2$ where $N$ is the
number of states in the composite system
and there are $d$ links with hopping integral $J_h$.  Since there are
$2NW/\Delta$ nonzero matrix elements, this
corresponds~\cite{Genway2012} to $c\sim d J_h^2/W$.  Unlike
in conventional statistical mechanics, we do \emph{not} assume a weak
system-bath coupling so that we can study local observables in a
homogeneous optical lattice. Such local measurements are becoming
experimentally accessible~\cite{Fukuhara2013}. 
Effects of time-reversal symmetry can be studied by trap rotation or
artificial gauge fields~\cite{Dalibard2011}.

{\it Central Result.}--- At time $t=0$, we prepare the total system in a
separable initial state $\ket{\Psi(0)}=\ket{S}\otimes\ket{B}$, for a
general system state $\ket{S}$.  The bath state $\ket{B}$ is
restricted only by the requirement that it should have a small energy
uncertainty.  This means $\ket{\Psi(0)}$ has significant overlap only
with eigenstates of $H$ centered around a total energy $E_0 =
\bra{\Psi(0)}H\ket{\Psi(0)}$. The system evolves as $\ket{\Psi(t)} =
e^{-iHt} \ket{\Psi(0)} = \sum_A e^{-iE_A
  t}\ket{A}\overlap{A}{\Psi(0)}$ where $\ket{A}$ are the exact
eigenstates of the composite system with energies
$E_A$ ($\hbar\,$=1).  We study the reduced density matrix (RDM)
obtained by tracing out the bath: $\rho_{ss'}(t) \equiv
\sum_b\overlap{sb}{\Psi(t)}\overlap{\Psi(t)}{s'b}$.  Our main result
is the full temporal evolution of the RDM in the limit of a large bath
($\Delta \ll c$, $W$), for times $t \ll 1/\Delta$:
\begin{align}
\rdmdiag(t) &\simeq \rdmdiag(\infty) + [\rdmdiag(0)-\rdmdiag(\infty)]
e^{-2\Lambda(0,t)}\,, \label{eq:rdmdiag_t}\\
\rho_{ss'}(t) &\simeq \rho_{ss'}(0) e^{-i(\egysys_s-\egysys_{s'})t}
e^{-2\Lambda(0,t)} 
\quad (s'\ne s) \,, \label{eq:rdmoffdiag_t} \\
\Lambda(t',t) &= 
\!\int_{-\infty}^\infty \frac{c(E) R(E)}{E^2} 
\left(e^{iEt'}- e^{iE(t-t')}\right)\; dE \,. \label{eq:Lambda}
\end{align}
Here, $c(E)$ is the profile for the banded coupling matrix: $c(E) = c$
for $|E|<W$ and zero otherwise.
The symmetries of the random matrix model enter via the level
repulsion~\cite{Mehta1967}, expressed by $R(E)\propto |E|$ or $E^2$
for systems with or without time-reversal symmetry, respectively, for
$|E|\lesssim\Delta$, and tending to unity for $|E|\gg \Delta$. 
Note that the thermalization dynamics discussed below is insensitive to time-reversal
symmetry because thermalization occurs over 
time scales shorter than the time scale $1/\Delta$ over which the system is
sensitive to level repulsion.

In this limit of a large bath, we find that the 
diagonal elements of the RDM decay to reach a steady-state value
expected from the Gibbs distribution $\rdmdiag(\infty) = \dosbath(E_0
- \egysys_s) \Delta$ where $\dosbath$ is the bath density of
states~\footnote{To be precise, $\Delta$ should be the average level
  spacing near total energy $E_0$.}. Moreover, decoherence has the
same dynamics as thermalization, with the off-diagonal elements
$\rho_{ss'}(t)$ tending to zero at long times on the same time
scales~\footnote{We note that decoherence and thermalization with an
  interacting bath can have different time scales if there are
  different couplings for elastic and inelastic processes.}.

Most importantly, we establish that the relaxation
towards the thermal state has two temporal regimes (as seen in our
numerics~\cite{Genway2010}). The RDM is controlled by $\Lambda(0,t) \simeq t^2
\int_{-\infty}^\infty\! c(E)dE = cWt^2$ for $t\ll W^{-1}$, and
$c(E\!\to\! 0)\pi t$ for $W^{-1}\ll t \ll\Delta^{-1}$. So, the RDM has a
Gaussian decay with a decay rate of $2\sqrt{c W}$ for $t<W^{-1}$ but has
an exponential tail at longer times with decay rate $2\pi c$. For weak
coupling ($c \ll W$), the decay is predominantly exponential, as
expected from Fermi's Golden Rule and perturbative Lindblad
theory. For stronger coupling~\footnote{We consider the limit of a
  large bath where $\Delta\to 0$ and the total bath bandwidth diverges
  at fixed $c$ and $W$. This does not cover the scenario (global
  quench) where $c$ diverges with $1/\Delta$ when the spectrum will be
  strongly modified.} ($c \gg W$), the Gaussian form dominates with
thermalization completed by the crossover time $W^{-1}$.
We stress that the existence of the Gaussian and exponential regimes
is robust as our results apply to a general $c(E)$, and the rates are
controlled by only two quantities: $\int_{-\infty}^\infty\!
c(E)dE\propto {\rm Tr}V^2$ and $c(E\to 0)$.

{\it Brownian Model.}--- We use the Dyson Brownian
technique~\cite{Dyson1962,*Dyson1972,Chalker1996a} which enables us to
calculate the ensemble-averaged effects of the random coupling $V$ by
building it up as a sum of uncorrelated random perturbations:
\begin{equation}\label{eq:brownian}
\Vcouple \to \Vbrown(\tau) = \int_0^{\tau}\!\! \xi(\tau') d\tau'
\mbox{\ with\ }\tau =1\,.
\end{equation}
It can be pictured as a random walk in fictitious time $\tau$ in the
space of random Hamiltonians.  At $\tau\!=\!0$ 
the system and bath are decoupled. Dyson observed that the
$\tau\!=\!1$ case corresponds, after ensemble averaging, to the model
defined in \eqref{eq:definemodel} with $H=H_0+V$. More precisely, at
each fictitious time step $\delta\tau$, a small perturbation
$\xi(\tau) \delta\tau$ is added to the Hamiltonian
$H(\tau)=H_0+\Vbrown(\tau)$ which has exact eigenstates
$\ket{A(\tau)}$. This perturbation can be written in the basis of
these eigenstates as $\bra{A(\tau)}\xi\,\delta\tau\ket{B(\tau)} =
\sqrt{c_{AB}}\xi_{AB}$. The banded coupling profile, defined after
Eq.~\eqref{eq:Lambda}, is mimicked by
$c_{AB} \equiv c(E)\Delta$, with $E=E_A-E_B$. (See the Discussion
for the validity of this approach.)    Restricting ourselves to
time-reversal-invariant systems, we model the randomness by the
independent Gaussian random variables $\xi_{AB}$ ($=\xi_{BA}$) with
the stochastic properties: $\expect{\xi_{AB}} = 0$, and $\expect{
  \xi_{AB}\xi_{CD}} = (\delta_{AC}\delta_{BD} +
\delta_{AD}\delta_{BC}) \delta\tau$. It can be
shown~\cite{Wilkinson1995} from perturbation theory that we have
Langevin processes for the eigenstates and eigenenergies:
\begin{align}
\delta \xlap{sb}{A} &= 
\sum_{B\ne A} \left(\frac{\sqrt{c_{AB}}\;\xi_{AB}}{E_{AB}}  \xlap{sb}{B} 
- \frac{c_{AB}\;\delta\tau}{2E_{AB}^2}\xlap{sb}{A}\right)\,,
\label{eq:langevin_wavefn}\\
\delta E_A& =
\sqrt{c_{AA}} \xi_{AA} +  
\sum_{B\ne A} \frac{c_{AB}\;\delta\tau}{E_{AB}}
\label{eq:langevin_energy}
\end{align}
where $E_{AB}\equiv E_A-E_B$, and the overlap
$\xlap{sb}{A}(\tau)\equiv\overlap{sb}{A(\tau)}$ is a component of the
eigenstate in the decoupled basis. The initial $(\tau=0)$ condition is
$\xlap{sb}{A}(0)=\overlap{sb}{A(0)}$. ($\ket{A(0)}$ is a product state
of system and bath eigenstates.)  The perturbations for the overlaps
and the energy levels involve independent (off-diagonal and diagonal)
elements of $\xi_{AB}$. So, we can replace the sum over energies in
\eqref{eq:langevin_wavefn} with statistical averages over the
well-known energy level distribution. Fluctuations should be small
owing to the rigidity of the spectrum.  The second moment of the
overlap, $\expect{|\xlap{sb}{A}|^2}$, is the `local density of states'
(LDOS) in energy space. Its Brownian motion has been
studied~\cite{Wilkinson1995} for an unbanded coupling matrix. We have
extended the theory to obtain the fourth moments of the overlap that
are needed for the RDM.

\emph{Derivation.}--- We will now describe our analytical calculation in
more detail.  We focus first on the diagonal elements of the RDM and
consider, for brevity, the case of an initial product state
$\ket{\Psi(0)} = \ket{s_0 b_0}$. (It is straightforward to generalize
to other initial product states.)  It is useful to express the RDM in
terms of the overlaps between the exact eigenstates and the decoupled
product states, $X_A^{sb}\equiv\overlap{sb}{A}$, which is of random
sign over the ensemble of random couplings. It can be shown that
\begin{equation}\label{eq:rdmdiag_def}
\rdmdiag(t) = \sum_{ABb}
 \overlap{A}{\sinit\binit}\overlap{\sinit\binit}{B}
 \overlap{B}{sb}\overlap{sb}{A} e^{-iE_{AB}t}\,.
\end{equation}
This involves the fourth moments of the overlaps. Let us start with
the second moments $J^{\alpha\beta}_A(\tau)\equiv
\expect{\xlap{\alpha}{A}\xlap{\beta}{A}}$ ($\alpha \equiv (ra)$,
$\beta\equiv (sb)$).  Using \eqref{eq:langevin_wavefn}, we can write
down the Langevin equation for
$\delta(\xlap{\alpha}{A}\xlap{\beta}{A})
=\xlap{\alpha}{A}\delta\xlap{\beta}{A}
+\xlap{\beta}{A}\delta\xlap{\alpha}{A}
+\delta\xlap{\alpha}{A}\delta\xlap{\beta}{A}$. Averaging over the
noise $\xi$ gives
\begin{equation}
  \partial_\tau J^{\alpha\beta}_A
  = \sum_{B\neq A}\frac{c_{AB}}{E^2_{AB}}
  \left(J^{\alpha\beta}_B-J^{\alpha\beta}_A\right)
\label{eq:eom_J}
\end{equation}
with $J^{ra,sb}_A(0) = \overlap{sb}{A(0)}\delta_{ra,sb}$. 
The sum is in the form of a convolution and so
this differential equation can be 
simplified in the time cdomain in terms of $J^{\alpha\beta}(t,\tau) =
\int J^{\alpha\beta}_A(\tau) e^{-iE_A t}dE_A/\Delta$. It
simply becomes $\partial_\tau J^{\alpha\beta}(t,\tau) =
\Lambda(0,t) J^{\alpha\beta}(t,\tau)$, and the solution is
\begin{equation}\label{eq:J_ttau}
J^{\alpha\beta}(t,\tau) = \delta_{\alpha\beta} e^{-iE_\beta t} e^{-\tau \Lambda(0,t)}\,.
\end{equation}
This is nonzero only if $\alpha=\beta$ because, upon averaging over
the random couplings $\Vbrown(\tau)$, there should be no correlations
between different components of $\ket{A}$ in the decoupled basis. (A
test is to consider the terms that survive under an average over random
gauge transformations of the set of the basis states $\ket{sb}$.)

We note that $J^{\beta\beta}(t,\tau\!=\!1)$ is the Fourier transform
of the local density of states $\expect{|\xlap{sb}{A}|^2}$. From the
behavior of $\Lambda(0,t)$ at short and long times as discussed after
\eqref{eq:Lambda}, we see~\footnote{This has been seen in our numerics~\cite{Genway2012}.} that the LDOS is a function of
$\omega= E_A-E_{sb}$ which is a Lorentzian of width $\pi c$ for
$\omega \ll W$, and is cut off at $\omega\gg W$ by a Gaussian of
width $\sqrt{2cW}$. 
The LDOS can also be obtained in large-$N$
diagrammatics for the random coupling where $N$ corresponds to the
number of bath states. For an unbanded matrix ($W\to \infty$), the
leading result corresponds to a self-consistent Born approximation,
giving the Lorentzian broadening to the
LDOS~\cite{LebowitzPastur2004}. However, the Gaussian tail for a
banded matrix is more difficult to capture in such an
approximation.

The result \eqref{eq:J_ttau} demonstrates analytically
ETH~\cite{Deutsch1991,*Srednicki1994,*Rigol2008} 
which gives a sum rule for the LDOS, \emph{i.e.}~the
projection of an eigenstate $A$ onto a system
state $s$:
$\sum_b\expect{|\overlap{A}{sb}|^2} = \sum_b J^{sb,sb}_A 
\propto \dosbath(E_A-\egysys_s)$. 
Using \eqref{eq:J_ttau}, we see that
\begin{equation}
\sum_b\expect{|\overlap{A}{sb} |^2} 
= \Delta\int_{-\infty}^\infty\!\! \dtwopi{t} e^{-\tau\Lambda(0,t)}
\sum_b e^{i(E_A-E_{sb})t} \,.
\end{equation}
This is a sum of the Fourier transform $r_1(\omega)$ of
$e^{-\tau\Lambda(0,t)}$ at frequencies $\omega= E_A-\egysys_s -
\egybath_b$ over all $\egybath_b$.  
Recall that $e^{-\tau\Lambda(0,t)}$ is mainly
Gaussian decay with a rate of $\sqrt{c\tau W}$ for $c\tau \gg W$ and
mainly exponential with rate $\pi c\tau$ for $c\tau \ll
W$.  Therefore, $r_1(\omega)$ should be a function
centered at $\omega=0$ with width
$\sim\textrm{min}[c\tau,\sqrt{c\tau W}]$. 
Assuming that the bath density of states
$\dosbath$ varies slowly over this width,
we find agreement with ETH:
\begin{align}\label{eq:eth}
\sum_b\expect{|\overlap{A}{sb}|^2} &\simeq 
\dosbath(E_A-\egysys_s) \Delta\int\!
r_1(E_A-\egysys_s-\egybath)d\egybath \notag\\
&\hspace{-4em}=  
\dosbath(E_A -\egysys_s) \Delta e^{-\tau\Lambda(0,0)} = \dosbath(E_A -\egysys_s) \Delta\,.
\end{align}

Let us now turn to the fourth moments needed for the evaluation of the
RDM, $M^{s\alpha}_{AB}(\tau)\equiv \sum_b \expect{\xlap{\alpha}{A}
  \xlap{\beta}{A} \xlap{\beta}{B} \xlap{\alpha}{B}}$ and
$N^{s\alpha}_{AB}(\tau) \equiv \sum_b [\expect{\xlap{\alpha}{A}
  \xlap{\alpha}{A} \xlap{\beta}{B} \xlap{\beta}{B} }+
(A\!\leftrightarrow\!B)]/2$, and their associated time-domain
functions $M^{s\alpha}(t,\tau) \equiv \int\!\!\int M_{AB}^{s\alpha}
e^{-iE_{AB}t} dE_AdE_B/\Delta^2$ and similarly for $N^{s\alpha}$.  It
can be shown~\footnote{See Section I of Supplemental Material.}
from the Langevin equations \eqref{eq:langevin_wavefn} that:
\begin{align} 
\label{eq:eom_mom4tM}
\hat D_\tau M^{s\alpha}(t,\tau)
&= 2\Delta\!\! \int\!\!\dtwopi{t'}
\Lambda(t',t) [M^{s\alpha}(t',\tau)\!+\! N^{s\alpha}(t',\tau)]\,,\\
\label{eq:eom_mom4tN}
\hat D_\tau N^{s\alpha}(t,\tau)
 &= 4 \Delta\!\!\int\!\!\dtwopi{t'}
\Lambda(t',t) M^{s\alpha}(t',\tau)
\end{align}
where $\hat D_\tau \equiv \partial_\tau + 2\Lambda(0,t)$ and 
the initial conditions are
$M^{s\alpha}(t,0)= \delta_{rs}$ and 
$N^{s\alpha}(t,0)= \sum_b\cos[(E_\alpha-E_{sb})t]$.
The latter sums over \emph{all} bath states. It is strongly
peaked at $t=0$ with a width of
the inverse bath bandwidth, and is approximately
$(2\pi/N_s\Delta)\delta(t)$. 

We will now proceed to a solution of these equations of motion for a large
bath ($\Delta\to 0$). Consider the $t'$-integrations over $M^{s\alpha}$ 
in the above equations. We can divide up
$M^{s\alpha}$ into its transient part and its steady-state value
$M^{s\alpha}_\infty(\tau)$ at long times. Anticipating
that the transient part decays exponentially at long times and does
not scale with $1/\Delta$ [see \eqref{eq:rdmdiag_t}], 
we expect that its contribution to the
integral should vanish with $\Delta$. The
contribution of the steady-state part $M^{s\alpha}_\infty(\tau)$
is proportional to
\begin{equation}
  M^{s\alpha}_\infty(\tau)\! \int_{-\infty}^\infty\!\!\!\!\!\Lambda(t',t)\,dt'
  \propto\int_{-\infty}^\infty\!\!\!\! \!dE f(E,t) R(E)\delta(E)
\end{equation}
where $f(E,t)=c(E)(1- e^{iEt})/E^2$.
This integral vanishes since $R(E) \sim |E|$ as $E\to 0$.
Hence, we find that $N$ is not coupled to $M$ in this
limit of $\Delta \to 0$ and that the solution to \eqref{eq:eom_mom4tN} is
simply $N^{s\alpha}(t,\tau) = e^{-2\tau \Lambda(0,t)} N^{s\alpha}(t,0)$.
Thus, the right side of 
\eqref{eq:eom_mom4tM} becomes
\begin{align}\label{eq:eom_simple}
\lefteqn{
\frac{\Delta}{\pi}\!\!\int\!\!dt'\, dE \sum_b 
f(E,t)e^{-2\tau\Lambda(0,t')-iEt'}\!\!
\cos\left[(E_{s\alpha}-\egybath_b)t'\right]
}\notag\\
&= \frac{\Delta}{2\pi}\!\!\int\!\!d\epsilon\, dE \, 
f(E,t) \dosbath(\epsilon)\!\!\sum_{\eta=\pm1}r_2(\egybath-E_{s\alpha}+\eta E)\,.
\end{align}
where $E_{s\alpha}=E_\alpha-\egysys_s$
and $r_2(\omega)$ is the Fourier transform of
$e^{-2\tau\Lambda(0,t)}$ which is peaked at zero with width
$\sim\textrm{min}[c\tau,\sqrt{c\tau W}]$. For a smooth $\nu_b$, 
$\dosbath(\egybath) \simeq
\dosbath(\egybath=E_{s\alpha})$ for the $\egybath$-range over which $r_2$
contributes to the $\egybath$-integration.  Then, the
right-hand side of \eqref{eq:eom_simple} becomes $2\Lambda(0,t)
\dosbath(E_{s\alpha})\Delta$. The equation of motion simplifies to
\begin{equation}\label{eq:eom_final}
[\partial_\tau+ 2\Lambda(0,t)]M^{s\alpha}(t,\tau) = 2\Lambda(0,t)
\dosbath(E_{s\alpha})\Delta \,.
\end{equation}
From \eqref{eq:rdmdiag_def}, $\rdmdiag = M^{s,\alpha =\sinit\binit}$
for an initial state $\ket{\sinit\binit}$ of energy $E_0 =
E_{s\alpha}+\egysys_s$.  The solution at $\tau=1$ for 
\eqref{eq:eom_final} is indeed our result
\eqref{eq:rdmdiag_t} with $\rho_{ss}(0) = \delta_{s\sinit}$.
We can perform an analogous calculation for $\rho_{s\ne s'}(t)$. The
dominant contributions~\footnote{See Section II
  of Supplemental Material.} come from
terms that are positive definite in the sum over bath states.
\begin{align}\label{eq:rdmoffdiag_approx}
\rho_{ss'}(t) &\simeq \rho_{ss'}(0)\sum_{AB}
 \expect{|\overlap{A}{s\binit}|^2 |\overlap{B}{s'\binit}|^2}
 \: e^{-iE_{AB}t}\notag\\
&\simeq \rho_{ss'}(0) J^{s\binit,s\binit}(t)J^{s'\binit,s'\binit}(-t)\,.
\end{align}
With \eqref{eq:J_ttau}, 
this gives our result \eqref{eq:rdmoffdiag_t} for decoherence.
\vspace{-\baselineskip}
\begin{figure}[bht]
  \includegraphics[scale=1.25]{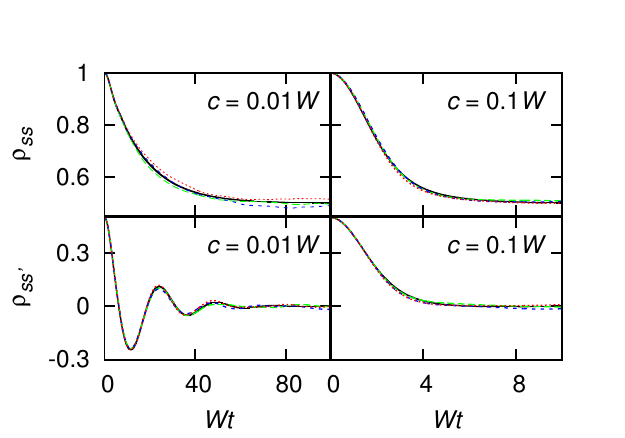}
  \caption{Comparison of Brownian motion result (solid) with
    exact diagonalisation (dotted) of 3 random realizations
    with 2 system states and 7000  bath
    states. Top: diagonal RDM elements, $\rho_{ss}$, for initial state
    $\ket{sa}$ with state $a$ near the center of the bath
    spectrum (high effective temperature).  Bottom: off-diagonal RDM
    elements, $\text{Re}(\rho_{ss'})$, for initial state
    $(\ket{sa}+\ket{s'a})/\sqrt{2}$. ($\egysys_s-\egysys_{s'}=W/4$,
    spacing $\Delta=W/4000$.) Weak
    coupling (left) shows predominantly exponential decay, while
    stronger coupling (right) shows the early Gaussian regime.}
  \label{fig:compare}
\end{figure}

\emph{Discussion.}--- Brownian motion
produces a random matrix, $\Vbrown(\tau=1)$, that has identical
statistical properties to $\Vcouple$ only for an unbanded random
matrix. For a banded coupling, this is only approximate. This is
because the coupling matrix is banded in the eigenstate basis of 
$H(\tau)$ at each Brownian step, instead of being banded
in the eigenbasis of $H(\tau=0)$. For finite $W$, we can show
~\cite{GenwayThesis} that $\Vbrown(\tau)$ has a broadened profile
$c'(E)$ for its matrix elements with increasing $\tau$.  As discussed
after \eqref{eq:Lambda}, the features of $c'(E)$ relevant to the
physics here are the integrated profile $\int c'(E)dE$ and small-$E$
limit of $c'(E)$. The former gives $\text{Tr}\Vbrown^2$ which has been
fixed at $\text{Tr}\Vcouple^2$ (implying that the Brownian model
reproduces the short-time expansion correctly: $e^{-iHt}\simeq 1 -
iHt$ giving $\rho_{\sinit\sinit} \simeq 1 - 2(cWt)^2$.) So, the
broadening of $c'(E)$ compared to $c(E)$ means that
$c'(0)<c(0)$. Thus, we overestimate the exponential decay rate,
but this is only significant when $c\gg W$ so that Gaussian
decay dominates and the exponential tail is negligible.  As we show in
Fig.~\ref{fig:compare}, our analytical results for
$H=H_0+\Vbrown(\tau\!=\!1)$ agree with the dynamics for $H=H_0 +
\Vcouple$.

To summarize, we have used a random matrix model to describe the
nonequilibrium dynamics of a system coupled to a fully
quantum-mechanical bath. In contrast with studies employing an
effective scattering approach~\cite{Dittes2000} with a non-Hermitian
random Hamiltonian, we study the full Hilbert space of a system with
an interacting quantum bath.  This provides an analytical
demonstration of the eigenstate thermalization hypothesis. (Many
previous works provided only numerical support.)  We also find that
thermalization and decoherence both follow the same dynamical
behavior, with Gaussian decay at short times and exponential decay at
long times.  We should point out that these two regimes have been
qualitatively anticipated in works based on semiclassical dynamics of
energy wavepackets~\cite{Hiller2006,*Cohen2000}. Also, a short-time
Gaussian regime was found~\cite{Flambaum2000,*Flambaum2001} for a
global quench that switches on a random two-body interaction among all
particles~\footnote{This corresponds to the parameter $c$ in this
  paper scaling with $1/\Delta$.}. That Gaussian decay originates from
the interactions generating a Gaussian density of states for the total
energy spectrum. In contrast, our \emph{local} quench does not alter
drastically the spectrum of the total system, and so we argue that the
Gaussian regime in our problem has a completely different physical
origin.  More recently, Gaussian decay has been found for a small
system coupled to a classical bath in a slow local quench
\footnote{L.~d'Alessio and A.~Polkovnikov, private communication.},
with a decay time controlled by the correlation time in the bath.  The
quench rate can be mimicked in our formalism by the width $W$.  Our
model has a short correlation time in the bath. Incorporating bath
correlations is the goal of future work.

We are grateful to John Chalker for useful discussions.  S.~G.~wishes to
thank the Leverhulme Trust (F/00114/B6) for financial support.

\bibliographystyle{apsrev4-1}
\bibliography{nanotherm}
\newpage

\begin{widetext}
\section{Supplemental Material}
\subsection{I. Diagonal elements of the reduced density matrix}
We consider here the fourth moments of the overlaps
$\xlap{sb}{A}(\tau)\equiv\overlap{sb}{A(\tau)}$ needed
for the evaluation of the reduced density matrix (RDM):
\begin{equation}
\momM^{\alpha\beta}_{AB}(\tau) \equiv 
\expect{\xlap{\alpha}{A} \xlap{\beta}{A}\xlap{\beta}{B}\xlap{\alpha}{B}}
\,,\qquad
\momN^{\alpha\beta}_{AB}(\tau) \equiv \frac{1}{2}
\left[\,\expect{\xlap{\alpha}{A} \xlap{\alpha}{A} \xlap{\beta}{B}
  \xlap{\beta}{B} }+ (A\leftrightarrow B)\right]
\end{equation}
where $\ket{\alpha}=\ket{ra}$ and $\ket{\beta}=\ket{sb}$ correspond to
system-bath product states. 

For a system prepared in the initial state 
$\alpha$, the quantities needed for the diagonal
elements of the RDM, $\rho_{ss}(\tau)$, are $M^{s\alpha}_{AB}(\tau) =
\sum_b \momM^{\alpha,\beta=sb}_{AB}(\tau)$ and $N^{s\alpha}_{AB}(\tau)
= \sum_b \momN^{\alpha,\beta=sb}_{AB}(\tau)$. 

Using the Langevin equations for the overlaps
and then averaging over the ensemble, we obtain two
coupled equations of motion for the Brownian motion:
\begin{align}\label{eq:eom_moments}
  \derivtau{\momM^{\alpha\beta}_{AB}} &=
  \sum_{D\neq A}\left[\green{AD}(\momM^{\alpha\beta}_{DB} - \momM^{\alpha\beta}_{AB}) + 
  (A\!\leftrightarrow\! B)\right]
  -2(1- \delta_{AB}) \frac{c_{AB}}{E^2_{AB}} 
  \left(\momM^{\alpha\beta}_{AB} + \momN^{\alpha\beta}_{AB}\right)
  + 2\delta_{AB}\!\! \sum_{D\neq A}\green{AD}
    \left(\momM^{\alpha\beta}_{AD} + \momN^{\alpha\beta}_{AD}\right)
  \notag\\
  \derivtau{\momN^{\alpha\beta}_{AB}} & =
  \sum_{D\neq A}\left[\green{AD} (\momN^{\alpha\beta}_{DB}- \momN^{\alpha\beta}_{AB})
    + (A\!\leftrightarrow\! B)\right]
  -4(1- \delta_{AB}) \frac{c_{AB}}{E^2_{AB}} \momM^{\alpha\beta}_{AB}
  + 4\delta_{AB} \sum_{D\neq A}\frac{c_{AD}}{E^2_{AD}} \momM^{\alpha\beta}_{AD}\,,
\end{align}
with $\momM^{\alpha\beta}_{AB}(0) = \delta_{\alpha\beta} \delta_{AB} \delta_{\alpha,A(0)}$
and $\momN^{\alpha\beta}_{AB}(0) =
(\delta_{\alpha,A(0)}\delta_{\beta,B(0)}+\delta_{\alpha,B(0)}\delta_{\beta,A(0)})/2$
where $A(0)$ and $B(0)$ are the decoupled states at $\tau=0$.

We are interested in the limit of $\Delta\to 0$ and so we will take
the continuum limit $\sum_D \to \int dE_D/\Delta$.  Since
$c_{AD}/E^2_{AD}$ is a function of only the energy difference
$E_{AD}=E_A-E_D$, we note that the terms involving sums over the exact
eigenstates $D$ are in the form of convolutions. Thus, the equation
can be simplified in the time domain. Let us define
$\momM^{\alpha\beta}(t,\tau) \equiv \int\!\!\int M_{AB}^{\alpha\beta}
e^{-iE_{AB}t} dE_AdE_B/\Delta^2$ and similarly for
$\momN^{\alpha\beta}$.  The equations of motion become:
\begin{align}\label{eq:eom_momM_t}
\left[\derivtau{} + 2\Lambda(0,t)\right] \momM^{\alpha\beta}(t,\tau)
&= 2\Delta\!\! \int\!\!\dtwopi{t'}
\Lambda(t',t) [\momM^{\alpha\beta}(t',\tau)\!+\! \momN^{\alpha\beta}(t',\tau)]\,,\\
\label{eq:eom_momN_t}
\left[\derivtau{} + 2\Lambda(0,t)\right] \momN^{\alpha\beta}(t,\tau)
 &= 4 \Delta\!\!\int\!\!\dtwopi{t'}
\Lambda(t',t) \momM^{\alpha\beta}(t',\tau)\,,
\end{align}
where $\Lambda(t',t)$ is defined by equation (4) in the main text.
The initial conditions at $\tau=0$ become
$\momM^{\alpha\beta}(t,0)= \delta_{\alpha\beta}$ and $\momN^{\alpha\beta}(t,0)=
\cos[(E_\alpha-E_\beta)t]$. 

These coupled equations of motion are linear equations in the moments.
So, setting $\beta=(sb)$ and summing over $b$, we see that
$M^{s\alpha}(t,\tau)$ and $N^{s\alpha}(t,\tau)$ [defined in the main
text above equation (13)] obey the same coupled set of equations as
$\momM^{\alpha\beta}(t,\tau)$ and $\momN^{\alpha\beta}(t,\tau)$.
Thus, we find the equations of motion in fictitious time $\tau$ for $M$ and $N$
as given in equations (13) and (14) of the main text.  The initial
conditions are $M^{\alpha\beta}(t,0)= \delta_{\alpha\beta}$ and
$N^{\alpha\beta}(t,0)= \sum_b\cos[(E_\alpha-E_{sb})t]$.

\subsection{II. Off-diagonal elements of the reduced density matrix}

Now, let us turn to the dynamics of decoherence. To observe decoherence,
we prepare the subsystem in an entangled state and then study the
off-diagonal elements of the RDM. Suppose we start with a single bath
state $a$, then
$\ket{\Psi(t=0)} = \sum_sc_s\ket{sa}$.
Then, the off-diagonal elements of the RDM are given by
($s\neq s'$)
\begin{equation}\label{eq:rhooffdiag}
 \rho_{ss'}(t) = \sum_{ABb}\sum_{rr'}
 \overlap{A}{ra}\rho_{rr'}(0)\overlap{r'a}{B}
 \overlap{B}{s'b}\overlap{sb}{A} e^{-iE_{AB}t} \,,\quad\text{with}\quad\rho_{rr'}(0)= c^*_{r'}c_r\,.
\end{equation}
The terms which survive averaging over the disorder should be the terms
which are invariant under random gauge transformation on the basis
states. 

Nonzero contributions come from the terms with $s=r$, $s'=r'$ and
$b=a$. These terms give
\begin{equation}\label{eq:rho1offdiag_N}
 \rho_{1ss'}(t) = \rho_{ss'}(0)\sum_{AB}
\expect{\xlap{sa}{A}\xlap{sa}{A}\xlap{s'a}{B}\xlap{s'a}{B}} e^{-iE_{AB}t}
= \rho_{ss'}(0) \momN^{sa,s'a}(t)
\end{equation}
Following an analogous analysis of the coupled equations (13) and (14)
of the main text, we expect that $\momN$ is decoupled from $\momM$ in
the thermodynamic limit ($\Delta\to 0$) so that \eqref{eq:eom_momN_t}
simplifies to $[\partial_\tau +
2\Lambda(0,t)]\momN^{sa,s'a}(t,\tau)=0$.  The solution is simply:
\begin{equation}
 \rho_{1ss'}(t,\tau) 
=  e^{-2\tau\Lambda(0,t)} e^{i(\egysys_s - \egysys_{s'})t}\rho_{ss'}(0)\,.
\label{eq:rho1offdiag}
\end{equation}
where $\egysys_s$ is the eigenenergy for subsystem state $s$.
This is the result given in equation (18) of the main text.

Another possible contribution to $\rho_{ss'}$ comes from setting
$s=r'$, $s'=r$ and $a=b$ in \eqref{eq:rhooffdiag}. This gives
\begin{equation}\label{eq:rho2offdiag_N}
 \rho_{2ss'}(t) = \rho_{s's}(0)\sum_{AB}
\expect{\xlap{s'a}{A}\xlap{sa}{A}\xlap{sa}{B}\xlap{s'a}{B}} e^{-iE_{AB}t}
= \rho_{s's}(0) \momM^{sa,s'a}(t)\,.
\end{equation}
$\momM^{sa,s'a}(t)$ obeys \eqref{eq:eom_momN_t} and is coupled to
$\momN^{sa,s'a}(t)$.  For $s\ne s'$, the initial conditions in
fictitious time are $M^{sa,s'a}(t,\tau =0) = 0$ and
$\bar{N}^{sa,s'a}(t,\tau =0) = \cos[(\egysys_s-\egysys_{s'})t].$ Since
$\momN$ is not of order $1/\Delta$ (unlike in our calculation for the
diagonal RDM elements), this does not contribute to the differential
equation for $\momM$ in the thermodynamic limit. This means that
\eqref{eq:eom_momM_t} simplifies to $[\partial_\tau +
2\Lambda(0,t)]\momM^{sa,s'a}(t,\tau)=0$. With the initial condition
that it is zero at $\tau=0$, this means that $\rho_{2ss'}=0$ in the
thermodynamic limit and does not contribute to the off-diagonal
elements of the RDM.

Note that the Langevin equations can be used to derive a 
Fokker-Planck equation for the joint
distribution $P(\{X\},\tau)$ for the overlaps:
\begin{equation}\label{eq:fokkerplanck}
\derivtau{P}=\!\!
\sum_{\substack{\alpha\beta AB\\B\ne A}} \!\frac{c_{AB}}{2E_{AB}^2}\!
 \left[\delta_{\alpha\beta}\frac{\partial (\xlap{\alpha}{A} P)}{\partial \xlap{\alpha}{A}}\right.
+ \frac{\partial^2 (\xlap{\alpha}{B}\xlap{\beta}{B} P)}{\partial \xlap{\alpha}{A}\partial\xlap{\beta}{A}}
- \left.\frac{\partial^2  (\xlap{\alpha}{B}\xlap{\beta}{A} P)}{\partial \xlap{\alpha}{A}\partial\xlap{\beta}{B}}\right]\,.
\end{equation}
All the results for the moments of the overlaps can be derived from
this Fokker-Planck equation.

\end{widetext}

\end{document}